# Photodissociation of water in crystalline ice:

# a molecular dynamics study

Stefan Andersson [a,b,*], Geert-Jan Kroes [b], Ewine F. van Dishoeck [a]

[a] *Leiden Observatory, P.O. Box 9513, 2300 RA Leiden, The Netherlands*
[b] *Leiden Institute of Chemistry, Leiden University, P.O. Box 9502, 2300 RA Leiden, The Netherlands*

**Abstract**

The photodissociation dynamics of a water molecule in crystalline ice at 10 K is studied computationally using classical molecular dynamics. Photodissociation in the first bilayer leads mainly to H atoms desorbing (65%), while in the third bilayer trapping of H and OH dominates (51%). The kinetic energy distribution of the desorbing H atoms is much broader than that for the corresponding gas-phase photodissociation. The H atoms on average move 11 Å before becoming trapped, while OH radicals typically move 2 Å. In accordance with experiments a blueshift of the absorption spectrum is obtained relative to gas-phase water.

[*] Corresponding author. Fax: +31 71 5274397.
*E-mail address*: s.andersson@chem.leidenuniv.nl (S. Andersson).

# 1. Introduction

Ultraviolet irradiation of ice is of great interest for understanding the chemistry in both atmospheric [1] and astrophysical environments [2]. In interstellar space, photodissociation of $H_2O$ molecules can be a driving force behind the chemistry on icy 'dust' grains in dense, cold (T = 10 K) molecular clouds even though the flux of UV photons is extremely low (~ $10^3$ photons $cm^{-2}$ $s^{-1}$) [2]. The mechanisms of such photoinduced processes are badly understood, however. For instance, it is not known if there is a significant probability of release of H atoms and OH radicals from the photodissociation of $H_2O$ molecules, since these fragments would be free to react with other species in the ice. Other possibilities include the recombination and / or direct desorption of H and OH. It is also interesting to understand the distribution of energy in the 'photofragments', since this will affect their reactivity. Finally, it is important to know how far the photofragments can travel upon photodissociation of the parent molecule.

A number of experiments have been performed on the UV absorption spectra of $H_2O$ ice [3], the formation of $H_2$ and other species [4-8], and the reconstruction (crystallization / amorphization) of $H_2O$ ice under UV irradiation [9-10]. There are theoretical studies of the excitation properties of liquid water [11-13] and water clusters [14,15], but there are no reported theoretical studies of the dynamics of photodissociation of a $H_2O$ molecule in ice. This paper is a first step in this direction, using a classical molecular dynamics (MD) approach with forces derived from analytical potentials.

## 2. Computational details

*2.1. The ice model*

To treat the photodissociation dynamics of a water molecule in ice, a method based on classical mechanics is used. A crystalline ice surface ($I_h$) is modeled as 8 bilayers of 60 $H_2O$ molecules each in a unit cell, using periodic boundary conditions in the *x* and *y* directions (in the surface plane). The unit cell is 22.4 Å by 23.5 Å in the *xy*-plane and the ice slab extends 29.3 Å in the *z* direction. The two lowest bilayers are kept fixed to simulate bulk ice and the motion of the remaining 360 molecules in the six moving bilayers is simulated using the molecular dynamics (MD) method [16]. The water molecules are treated as rigid bodies, with the exception of one molecule that is chosen to be dissociated (see Section 2.2). The three atoms of the dissociating molecule are able to move without dynamical constraints. Note that the rigidity of the molecules excludes the possibility of vibrational energy transfer to intramolecular vibrational modes and reactions involving bond breaking in $H_2O$ molecules not initially photodissociated. Nor can the 'solvated electron' observed after UV irradiation of liquid water and ice [17-20] be treated explicitly. The initial ice configuration obeys the ice rules [21] and has a zero dipole moment. The ice surface was prepared by equilibrating the motion of the water molecules over a period of 100 ps. The basic methodology of simulating the ice surface has been described in Ref. [22].

*2.2. Potentials*

The potentials used in this work are of three main types: (i) $H_2O$-$H_2O$ potentials, (ii) H-$H_2O$ and OH-$H_2O$ potentials, and (iii) the intramolecular potential for the dissociating $H_2O$ molecule. The $H_2O$-$H_2O$ potential used in constructing the ice surface is the TIP4P pair potential [23], which consists of an O-O Lennard-Jones (LJ) potential and electrostatic interactions based on charges on the H atoms and an additional charge site M (M: -1.04 $e$; H: 0.52 $e$). To avoid the difficulties of having a massless charge site in a fully flexible molecule, the TIP3P model [23] is used as a starting point for describing the intermolecular interactions of the dissociating molecule. The TIP3P potential consists of charges situated on the atoms (O: -0.834 $e$; H: 0.417 $e$) and an O-O LJ potential. To treat the intermolecular interactions of the excited $H_2O$ molecule, new atomic charges are used (O: 0.4 $e$; H: -0.2 $e$), which were set to reproduce the dipole moment of the gas-phase excited-state $H_2O$ molecule [24], with the TIP3P LJ potential kept the same. A similar approach was recently taken by Miller et al. [14] in their work on the excitation spectra of water clusters. Note that during the equilibration of the ice surface all intermolecular potentials are described using TIP4P, and only at the onset of the photodissociation calculation (see Section 2.3) is the intermolecular interaction of the photoexcited $H_2O$ changed.

The H-$H_2O$ potential is a reparameterization of the YZCL2 $H_3O$ potential energy surface (PES) [25], using an expression consisting of atom-atom repulsive and dispersion interactions and a H-O Morse potential [26]. Comparing with the previously available H-$H_2O$ potential [27], the long-range attractive part is roughly the same, but the short-range repulsive interaction is clearly improved [26]. The OH-$H_2O$ potential was obtained by fitting *ab initio* post-Hartree-Fock MP2 energies to a pair potential model modelling electrostatic interactions based on atomic charges that are dependent on the OH bond length, repulsive interactions, and dispersion interactions [26].

To switch between the potentials in different bonding situations (HOH and H+OH), switching functions were devised, which depend on the OH distances in the dissociating molecule. These functions are different from one or zero only in a limited region and smoothly connect two regions of different bonding situations with each other [26].

The intramolecular interactions in the dissociating molecule are governed by the potential energy surfaces developed by Dobbyn and Knowles (DK) for the ground and first excited state of the gas-phase $H_2O$ molecule [28]. Initially the system is on the first excited state PES, but when one of the OH bonds is between 3 Å and 3.5 Å, a switch is made to the ground state PES (along with the corresponding intermolecular potentials) [26]. In the switching region the potential is a linear combination of the excited-state and ground-state potentials. The choice of distances for the switching region is based on the fact that this is where the PESs become near degenerate. The reason for this is that both ground state and first excited state correlate with $H(^2S) + OH(^2\Pi)$ asymptotically. This switch can therefore be made without introducing troublesome 'kinks' in the potential. The procedure guarantees conservation of total energy since there is a smooth transition between the PESs. Should the stretching of the OH bond be reversed before it reaches 3.5 Å, the system will be on the excited state potential if the OH bond becomes shorter than 3 Å again. Otherwise, the dynamics is thereafter solely governed by the ground state PES. The justification of this admittedly *ad hoc* procedure is based on the assumptions that the probability of electronic deexcitation due to collision-induced internal conversion is high and that upon reencounters of H and OH the system is most likely to be on the ground-state PES. This allows for the possibility of recombination. The amount of recombination found in our calculations is expected to form an upper bound to the 'real' recombination probability, since reencounters are in reality not guaranteed to occur on the ground-state PES. The intermolecular potential of the recombined $H_2O$ molecule is described using TIP3P.

To avoid interaction of molecules with their periodic images, the interaction of molecules more than 10 Å apart are switched off [26]. This was not done for the internal interaction of the dissociating molecule, however, allowing for recombination of H (OH) with a periodic image of OH (H).

*2.3. The photodissociation model*

Following equilibration of the ice, one molecule was chosen to be photodissociated. Only molecules in the top three bilayers were considered. This is because photodissociation in the lower bilayers would be more likely to lead to fragments leaving the surface through the rigid 'bottom' than through the non-rigid top of the surface.

There are four distinct orientations of the molecules in a bilayer and three molecules of each orientation were picked per bilayer. Thus 36 molecules in total, 12 per bilayer, were chosen to be photodissociated to get a representative sample of starting configurations. Following the approach of van Harrevelt et al. [29] for gas-phase $H_2O$ photodissociation, a Wigner (semiclassical) distribution function was fitted to the ground-state vibrational wavefunction of the $H_2O$ molecule as calculated using the DK PES. From the Wigner distribution the initial coordinates and momenta of the atoms in the dissociating molecule are sampled using a Monte Carlo procedure. Note that the gas-phase vibrational wavefunction is used as a basis for calculations in the condensed phase. This is assumed to be a minor approximation, however. The excitation is taken to be of a Franck-Condon type and therefore the atoms are given the coordinates and momenta taken from the ground state as described above, their subsequent motion being governed by the *excited state* potential energy surface (PES). Note that this procedure gives an initial phase-space distribution of the dissociating molecule based on quantum mechanics. This has been shown to give good agreement between classical and quantum treatments of gas-phase photodissociation [30].

*2.4. Trajectories*

For each of the 36 molecules chosen, 200 trajectories were run. The time step was 0.02 fs and the maximum propagation time for the trajectories was set to 20 ps. The trajectories were terminated if the three atoms of the dissociating molecule obeyed any of the following criteria: (i) the atom was more than 10 Å above the surface (in the positive z direction) (desorption), (ii) the atom was at a negative z (below the ice slab), or (iii) the kinetic energy of the atom was smaller than the absolute value of the intermolecular potential it experienced (given the potential was negative) (trapping). Note that the O and H atoms in OH or $H_2O$ are treated individually in this scheme. The potential evaluated in (iii) does not include the intramolecular potential, but only the intermolecular part of the potential, which governs the binding of the molecules within the ice. Therefore, also a highly vibrationally excited OH or $H_2O$ can be considered to be trapped when the molecule is at the outer turning points of its vibrational motion, i.e. when the vibrational energy is very low. The gas-phase results reported in Section 3 have been calculated by sampling 1000 initial configurations in the same way as described in Section 2.3 and then solving the classical equations of motion describing the dissociation of the $H_2O$ monomer on the excited state PES.

*2.5. Absorption spectrum*

The absorption spectra for each bilayer that are discussed in Section 3.4 were calculated by first taking the energy difference between the excited and ground state potentials (including intramolecular and intermolecular potentials) for 1000 configurations of the excited $H_2O$ for the 12 molecules chosen as described in Section 2.3. The gas-phase spectrum is calculated in the same way using 1000 configurations but with the intermolecular interactions turned off. The transition dipole moment was taken to be independent of the geometry of the water molecule and the absorption spectrum is calculated as the number of times excitation energies occur within 0.1 eV wide energy intervals.

**3. Results and discussion**

*3.1. Direct desorption vs. trapping*

In Figure 1 the outcome of the trajectories is plotted in a bar diagram, presenting results for each bilayer. The first set of bars gives the probability of the H atom desorbing from the surface with the OH becoming trapped on or in the ice. The second set shows the probability of H and OH being trapped within or on top of the ice and the third one gives the fraction of trajectories where H and OH recombine and remain trapped as $H_2O$. In the fourth set the remaining outcomes are collected together. In the first bilayer several other outcomes are possible: (i) the desorption of OH leaving the H atom trapped, (ii) desorption of both H and OH, and (iii) desorption of the recombined $H_2O$. For the rest, this last set contains events that are artefacts due to the adopted model. These include penetration of the H atom through the bottom of the ice slab and recombination of the H atom with a periodic image of the OH radical. In 3 out of the 7200 trajectories $O(^1D)$ and $H_2$ were formed from recombination of OH and H and subsequent redissociation. The DK PES gives a good description of $O + H_2$ configurations so this is no artefact. Trajectories that were not terminated after 20 ps also belong to the 'other' category. They constitute less than 1% of the total number of trajectories.

Water molecules photodissociated in the first bilayer lead mainly to H atom desorption (65%) with a smaller fraction (17%) giving trapping of H and OH. In the second bilayer these two outcomes are roughly of equal importance (34% vs. 39%), while in the third bilayer trapping of H and OH dominates (51%) and in only 17% of the cases an H atom desorbs. It is interesting to note that the fraction of H atoms desorbing decreases by about 50% per bilayer going deeper into the ice. Recombination of H and OH and subsequential trapping of the recombined molecules is insignificant in the first bilayer (3%), but due to caging effects this fraction rises to 21% for molecules dissociated in the third bilayer. In a model with flexible $H_2O$ molecules in the ice, the fraction of recombined $H_2O$ could possibly increase since $H_2O$-$H_2O$ vibrational energy transfer has been shown to be very efficient in ice and liquid water [31-32]. Transfer of the excess vibrational energy of the recombined $H_2O$ to the intramolecular vibrational modes of the surrounding $H_2O$ molecules would lower the rate of redissociation (into H and OH) and therefore stabilization of the recombined molecule could be more efficient than in the present simulations.

*3.2. Mobility of the photofragments*

The mobility of H and OH has been monitored by calculating the distances of the atoms from their original positions at $t = 0$ (see Figure 2). The results for the first bilayer (Fig. 2a) and the third bilayer (Fig. 2b) are quite similar and show that the H atoms can travel over significant distances (up to 50 Å). Motion over the surface facilitates the motion for the H atoms originating in the first bilayer. The peak and average distances for the first bilayer are 10 Å and 12 Å, while for the third bilayer these distances are 6 Å and 11 Å, respectively.

The OH fragments are in general not as mobile as the H atoms. The average distance travelled for OH originally in the first bilayer is only 2.7 Å, but OH has been found to move over the surface by up to 70 Å. The reason for the higher mobility compared to the H atoms is that the OH radical has a much stronger attractive interaction with the ice surface than H does. Thus, whereas H atoms moving over the surface will most likely desorb quickly, OH can keep moving parallel to the surface for much longer times. OH radicals from the third bilayer are of course more restricted by the surrounding molecules and move on average only 1.9 Å with a maximum distance of 5 Å.

The mobility of the recombined $H_2O$ molecules is important for understanding the reconstruction of the ice during UV irradiation. For molecules from the first bilayer the average distance of the recombined $H_2O$ from the position of the 'original' $H_2O$ is 2.8 Å. For the third bilayer the average distance is 1.9 Å. This clearly suggests the possibility of reconstruction and phase transformation of UV irradiated ice, as has been observed by for instance Leto and Baratta [10].

*3.3. Energy distributions*

In Figures 3 and 4 the energetics of the released fragments are presented. Figure 3 shows the kinetic energies of the H atoms released following gas-phase photodissociation of $H_2O$ and the H atoms that eventually desorb following photodissociation of $H_2O$ molecules in the first and third bilayers. Results for the second bilayer are similar to those of the third bilayer. From Fig. 3 it is seen that the kinetic energy distribution for photodissociation in ice is broader than that from gas-phase photodissociation. For photodissociation of water in the first bilayer, there is a peak intensity of the released H atoms around 3.2 eV (compared to 1.8 eV for gas-phase photodissociation), but the energy distribution has significant populations down towards zero kinetic energy and the average energy is 2.1 eV (gas phase: 2.0 eV). The high-energy components correspond mostly to atoms that are released directly to the gas phase without colliding with the surrounding $H_2O$ molecules, while the lower-energy H atoms have lost part of their kinetic energy to the surroundings. It can be seen that the kinetic energy distribution for the third bilayer peaks around 0.2 eV, the average kinetic energy being 1.0 eV. This shows that energy loss due to inelastic collisions with the surrounding water molecules is almost inevitable for H atoms released following dissociation of $H_2O$ molecules belonging to the lower bilayers.

Figure 4 compares the vibrational state distribution of trapped OH molecules to that of OH following gas-phase photodissociation. This distribution was computed by applying box quantization to energy levels taken from a gas-phase Morse potential fitted to experiment [33]. It is clearly seen that the OH vibrational state distribution is colder for the ice photodissociation than for the gas-phase counterpart, even though the excitation energies are lower in the latter case (see Section 3.4). Differences between the bilayers are small and the average OH vibrational energy is 0.4 eV for dissociation of $H_2O$ in all three bilayers, compared to 0.6 eV for the gas phase. This difference in vibrational energy is most likely connected to the repulsive intermolecular interactions, which the excited state $H_2O$ experiences. This will initially 'slow down' the OH vibrational motion. It therefore seems that the 'extra' excitation energy experienced by the condensed-phase molecules compared to the gas phase goes into translational energy of the H atoms, rather than increased vibrational energy of OH.

The vibrational deexcitation of OH following $H_2O$ dissociation is insignificant on the time scale of our trajectories. This is because the OH vibration is only weakly coupled to the phonon modes of ice. If the $H_2O$ molecules in the ice were allowed to be flexible, vibrational relaxation might be faster due to near-resonant intermolecular vibrational energy transfer.

*3.4. Absorption spectrum*

The potentials adopted give a peak excitation energy of 9.3 eV for $H_2O$ molecules in the first bilayer and 9.6 eV for molecules in the second and third bilayers as seen in Figure 5. This is a considerable blueshift from the gas-phase peak, which is found at 7.4-7.5 eV both experimentally [34] and theoretically [35]. The experimental value of the peak crystalline ice excitation energy (at 80 K) is 8.6 eV [3] (Figure 5). Even though there is an apparent deviation of about 1 eV from experiment, the shape of the first absorption band is reasonably well reproduced, taking the third bilayer as representative for the bulk. This suggests that it should be possible to obtain better agreement with experiment by modifying the intermolecular potential. The use of the gas-phase dipole moment in determining the charges for the excited state molecule might not be a good approximation. It has been used here in view of the lack of reliable estimates of the condensed-phase effective dipole moment of the molecule in the first excited state. Due to its very high dipole polarizability [24], the electronically excited molecule will however most likely have a smaller or even reversed effective dipole moment in the condensed phase (see e.g. Ref. [11]). The Rydberg character of this state also gives a very delocalized charge distribution, which could lead to a significant exchange repulsion interaction with neighboring molecules. There will also be an effect on the intramolecular potential, but that is assumed to be minor compared to the intermolecular part. A refined model for the excited state inter- and intramolecular interactions is currently being developed.

## 4. Conclusions

A theoretical model of the dynamics of photodissociation of crystalline ice has been explored. Even though it is a simplistic model based on classical dynamics and analytic potentials, it gives valuable insight into the dynamics of photodissociation in the solid phase.

The desorption of H atoms formed in the photodissociation of $H_2O$ strongly depends on the bilayer to which the parent molecule belonged, with 65% of the H atoms desorbing from the first bilayer but only 17% from the third bilayer. The distribution of kinetic energy of the H atoms desorbing from the ice surface is much broader than the corresponding distribution from gas-phase $H_2O$ photodissociation. Whereas H atoms originating in the first bilayer show a peak at 3.2 eV, it shifts to 0.2 eV for the second and third bilayers, indicating a considerable loss of kinetic energy to the surrounding molecules. H atoms are also able to move considerable distances through or over the ice (up to 50 Å). The OH fragments are in general less mobile, with the exception of those originating in the first bilayer which can move up to 70 Å. This opens up the possibility of reaction of H and OH with species not in the absolute vicinity of where photodissociation took place.The OH vibrational state distribution from ice photodissociation is predicted to be colder than that from the corresponding gas-phase photodissociation.

The theoretical absorption spectrum of water in ice shows a blueshift of 2 eV with respect to the gas phase spectrum compared to an experimental value of about 1 eV. This suggests that improvements in the model are needed. Loss of energy to intramolecular vibrational modes of the surrounding $H_2O$ molecules and reactive interactions of H and OH with $H_2O$ molecules may also change results somewhat. Improvements of our model will be explored in future research.

**Acknowledgements**

We are grateful to Dr. R. van Harrevelt for making his code for gas-phase photodissociation available and to Dr. A. Al-Halabi and Dr. L. Valenzano for valuable discussions and contributions to the potentials used. Funding for this project was provided by a NWO Spinoza grant and a NWO-CW programme grant.

**Figure captions**

Figure 1. Outcomes of photodissociation of $H_2O$ in the top three bilayers of ice. Error bars correspond to 95% confidence intervals.

Figure 2. Distribution of the distances of trapped H atoms from their position at $t = 0$ in the first and the third bilayers. Error bars correspond to 95% confidence intervals.

Figure 3. Kinetic energy distributions of H atoms desorbing from the ice surface for photodissociation in the gas phase, and in the first bilayer, and the third bilayers of ice. Error bars correspond to 95% confidence intervals.

Figure 4. OH vibrational state distribution obtained using a box quantization procedure, for photodissociation of gas phase $H_2O$ and for photodissociation of $H_2O$ in the first three bilayers of ice.

Figure 5. Calculated absorption spectra for the gas phase and the top three bilayers and part of an experimental absorption spectrum of crystalline ($I_h$) ice at 80 K (Ref. [3]). (Color)

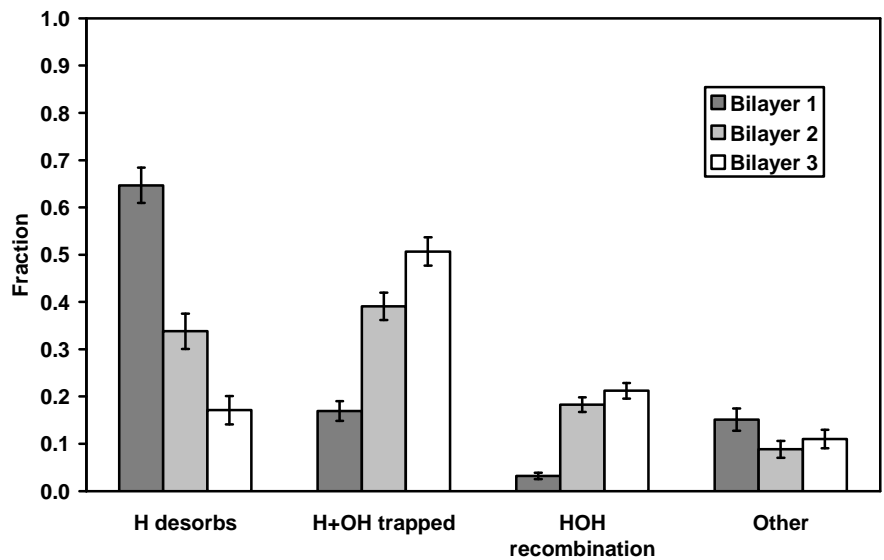

**Figure 1.** Andersson et al.

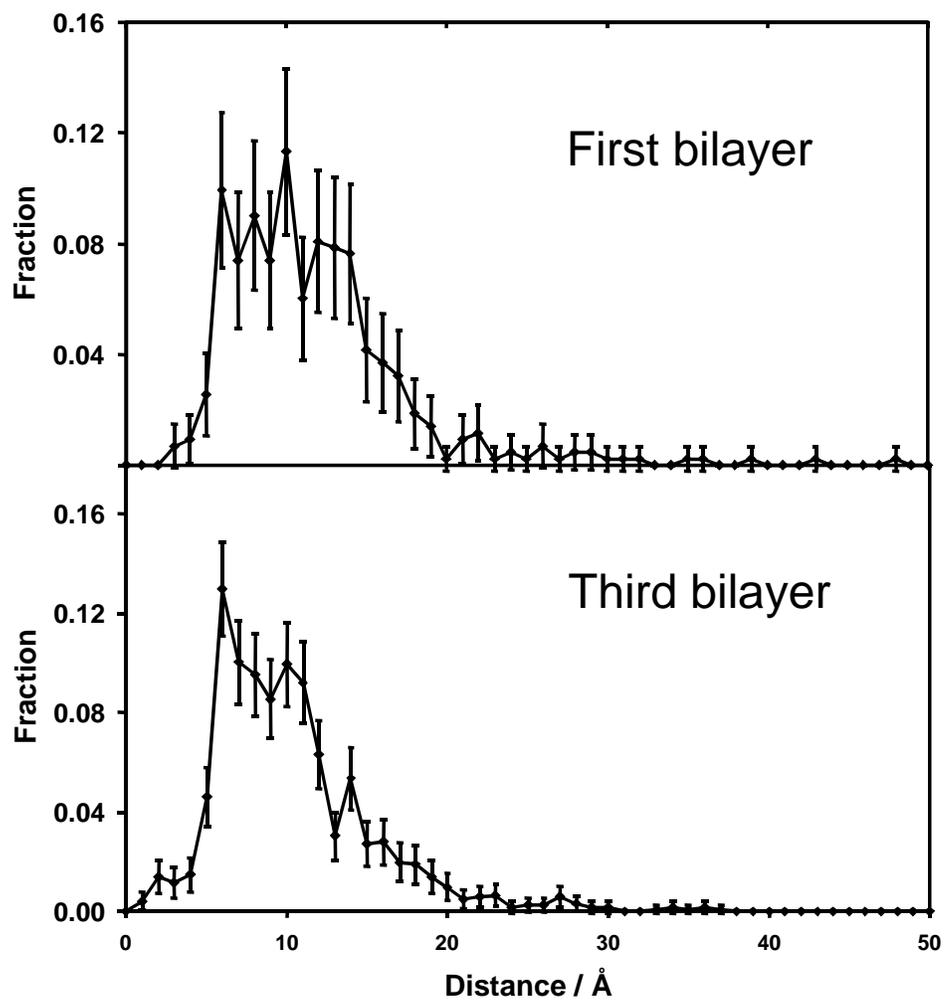

**Figure 2.** Andersson et al.

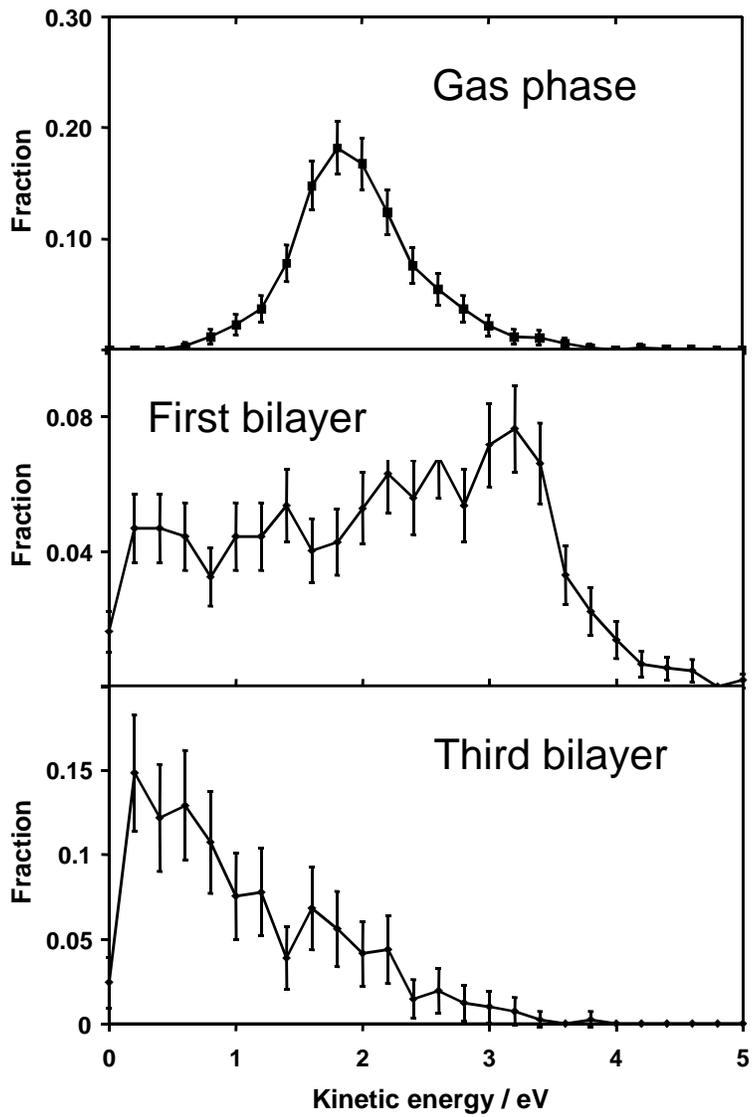

**Figure 3.** Andersson et al.

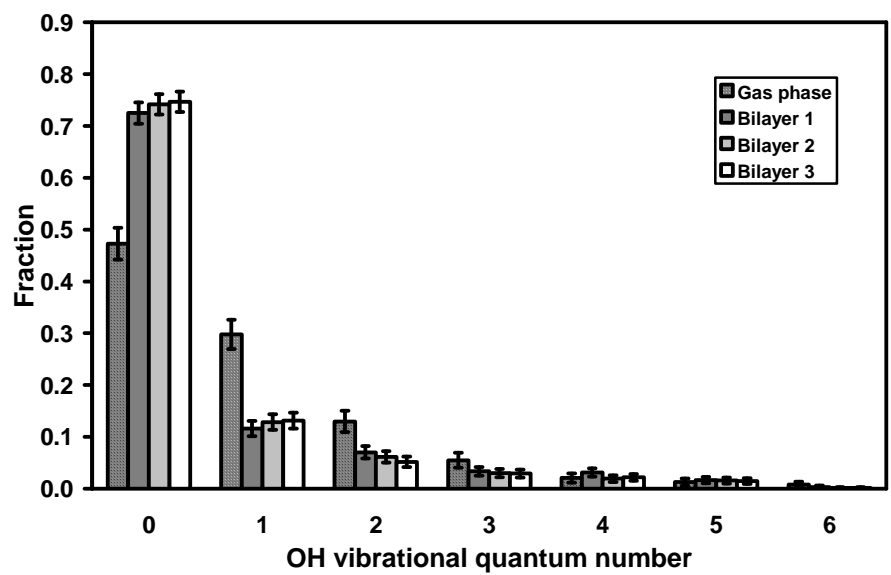

**Figure 4.** Andersson et al.

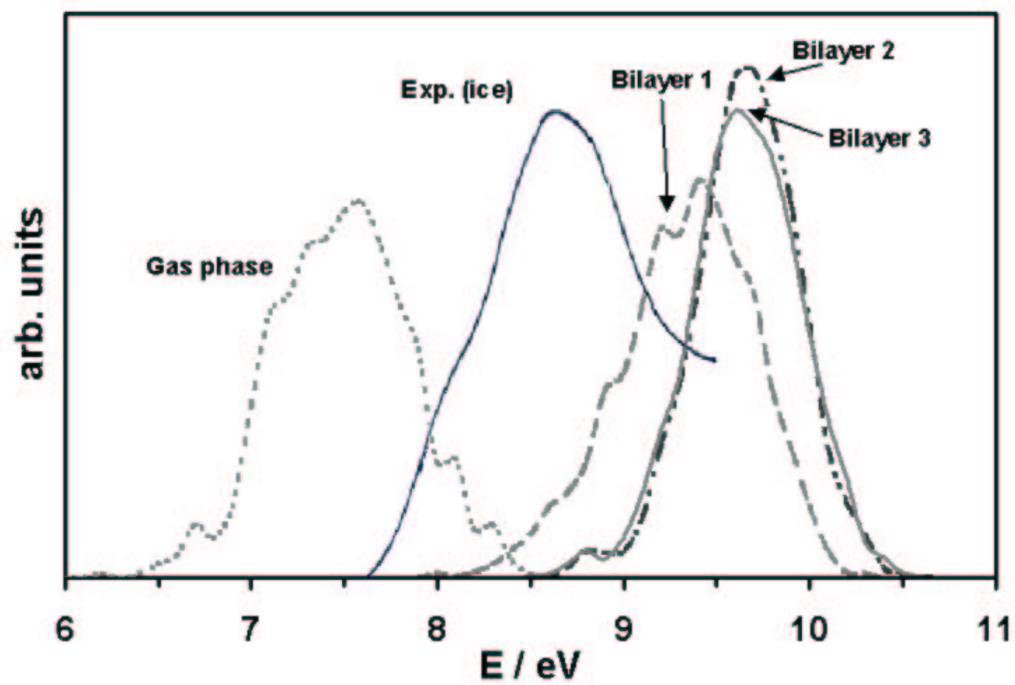

**Figure 5.** Andersson et al.